\documentclass{llncs}
\let\llncssubparagraph\subparagraph
\let\subparagraph\paragraph
\usepackage[compact]{titlesec}
\let\subparagraph\llncssubparagraph
\usepackage{mathptmx}
\usepackage{graphicx}
\usepackage{url}
\usepackage{color}
\usepackage{cite}
\usepackage[font={small,it}]{caption}
\usepackage{hyperref}
\usepackage[inline]{enumitem}
\usepackage{multirow}
\usepackage{wrapfig}
\usepackage{lipsum}

\newcommand{\myparagraph}[1]{\noindent{\bf #1}}

\title{Short Paper: On Deployment of DNS-based Security Enhancements}
\author{Pawel Szalachowski and Adrian Perrig}
\institute{ETH Zurich, Switzerland}
\date{}

\begin{document}
\maketitle

\begin{abstract}
    Although the Domain Name System (DNS) was designed as a naming system, its
    features have made it appealing to repurpose it for the deployment of novel
    systems.  One important class of such systems are security enhancements, and
    this work sheds light on their deployment.  We show the characteristics of
    these solutions and measure reliability of DNS in these applications.  We
    investigate the compatibility of these solutions with the Tor network,
    signal necessary changes, and report on surprising drawbacks in Tor's DNS
    resolution.
\end{abstract}

\section{Introduction}
\label{sec:intro}
DNS is one of the most successful Internet infrastructures.  It is a naming
system for resources over the Internet, and its most prominent use is to
translate human-readable names to IP addresses.  Currently, this hierarchical
and distributed system is a core infrastructure of the Internet, and over the
years the availability and reliability of standard DNS operations have
increased~\cite{pappas2004impact}.  Although DNS is primarily (and was designed
as) a system for name resolution, due to its success and flexibility it is used
by various, not initially intended, applications.  One family of such
applications are various security enhancements. These systems are particularly
difficult to deploy~\cite{nikkhah2015didn}, as different actors are reluctant to
deploy and invest in a security-dedicated infrastructure.  Due to low cost,
well-understood operations and administration, and its ubiquity, DNS seems like
an ideal environment to support deployment of new security enhancements.  Thus, it is naturally
appealing to protocol designers to repurpose the DNS infrastructure, rather than
designing and deploying a new one.  For those reasons, DNS is currently being employed
by various security enhancements.  As a consequence, new systems rely on the
infrastructure designed decades ago.  Therefore, it is necessary to investigate
how robust and applicable the infrastructure is for these use cases.  The
essence of the new uses is to transport additional information using DNS,
however, there exist indications that such a transport can be unreliable.

In this work we make the following contributions:
\begin{enumerate*}[label=\textbf{\textit{\arabic*}})]
    \item investigate the use of DNS-based security enhancements,
    \item study DNS reliability for these applications,
    \item check the compatibility of the enhancements if the DNS resolution
        occurs over Tor.
\end{enumerate*}

\section{Background}
\label{sec:pre}
\myparagraph{DNS Resolution} is a process of translating human-readable domain
names to IP addresses.  It is conducted through the DNS infrastructure, namely
\textit{DNS clients}, \textit{resolvers}, and \textit{servers}.  To resolve a
domain name, e.g., \texttt{www.a.com}, a client initiates the process by
querying its resolver, which in turn contacts one of the \textit{DNS root
servers} (root servers' IP addresses are fixed and known to resolvers).  The
root server returns an address of a \textit{DNS authoritative server} for the
\texttt{com} domain. Then, the resolver queries the \texttt{com} authoritative
server to find an authoritative server for \texttt{a.com}, which finally is
queried about \texttt{www.a.com}. The \texttt{a.com} authoritative server
returns the IP address(es) of \texttt{www.a.com}.  The lengthy resolution
process is usually shortcut by using cached information.

DNS allows to associate various information with domain names. Information is
encoded and delivered within \textit{resource records} (RRs) with dedicated
types, e.g., \texttt{A} and \texttt{AAAA} RRs map domain names to IPv4 and IPv6
addresses, respectively, \texttt{NS} RRs indicate authoritative servers, while
\texttt{TXT} RRs can associate an arbitrary text. DNS responses can contain
multiple RRs of the queried type.  It is also possible to translate IP addresses
into domain names (to this end \texttt{PTR} RRs are used). 

DNS deploys UDP as a default transport protocol, however, for responses larger
than 512 bytes a \textit{failover} mechanism is introduced.  Larger responses
are truncated to fit 512 bytes and marked by a \texttt{truncated} flag.
Resolvers receiving a truncated response query the server again via TCP to
obtain the complete response.  (Clients can increase the limit by signaling the
maximum UDP response size they can handle~\cite{rfc2671}.)

\myparagraph{DNS Resolution (Un)Reliability.}
Although DNS is reliable for its major application (i.e., translating names to
IP addresses), the reliability for other applications is questionable.  For
instance, many of DNS clients, resolvers, and servers are realized as
non-compliant implementations~\cite{dnsissues}.  It was
reported~\cite{huston2013question} that a significant fraction of all clients
(2.6\%) and a large fraction of resolvers (17\%) cannot perform the UDP-to-TCP
failover.  This behavior limits clients ability to receive responses larger than
512 bytes.    Another potential issue~\cite{dnsissues} is caused by network
environments, where devices can handle only unusually small Maximum Transmission
Unit (MTU) packets, thus introducing IP fragmentation decreasing the reliability
of the DNS resolution.  DNS traffic is also a subject to traffic analysis, and
some middleboxes manipulate DNS
responses~\cite{hatonen2010experimental,weaver2011implications}.  It is believed
that some non-standard RRs are discriminated by non-compliant implementations
or/and network devices. For instance, some
experiments~\cite{weaver2011implications,notdane} indicate that \texttt{A} RRs
are more reliable than \texttt{TXT} RRs.

\section{Security Enhancements Employing DNS}
\label{sec:enhancements}
We focus our study on two families of security enhancements that can benefit
from a robust DNS infrastructure, namely email and TLS PKI enhancements.  The
main reason why DNS infrastructure can be appealing for these technologies is
that both email and TLS PKI are domain based. As the DNS lookup usually precedes
the email exchange or TLS connection establishment, the client can obtain some
relevant information before the connection setup. Additionally, such DNS-based
information pre-fetching does not violate the privacy, as no additional third
party is contacted (DNS servers are contacted anyways).  A security assumption
for these schemes is that an adversary cannot control DNS entries of targeted
domains.

\subsection{Email}
\myparagraph{SPF}~\cite{rfc7208} enables domains to make assertions (in DNS)
about hosts that are authorized to originate email for that domain.  When an email
is received by an email exchanger, it parses the domain name from the email's
\texttt{From} address field, and queries the DNS to check whether the sender is
authorized to send email.  This mitigates spam and phishing emails that abuse
the \texttt{From} field.    SPF mainly uses \texttt{TXT} RRs, although a
dedicated \texttt{SPF} RR was introduced. 

\myparagraph{Sender ID}~\cite{rfc4405} is an anti-spoofing proposal based on
SPF.  The main difference is that it aims in verifying the sender address
displayed to an email client (the \texttt{From} field and the address displayed
by email clients can differ).  Such an address is introduced as a Purported
Responsible Address (PRA)~\cite{rfc4407}.  By setting a special \texttt{TXT} or
\texttt{SPF} record, a domain can specify if only SPF should be verified,
or both SPF and PRA, or PRA only.

\myparagraph{DKIM}~\cite{rfc6376} is an email authentication protocol based on
signatures.  A domain publishes RR with its public key.  Next, the domain's
outbound email server signs sent emails with the corresponding private key.  An
inbound email server, after receiving a signed email, extracts its origin domain
name (via the \texttt{From} field) and performs a DNS lookup to obtain the
domain's public key used to verify the email.  Usually, DKIM is executed by
email servers rather than email clients (i.e., authors and recipients). Public
keys are stored in \texttt{TXT} RRs, and to obtain a key of \texttt{a.com},
\texttt{\_dkim.a.com} is queried.  DKIM protects emails from modification,
however, the scheme can be bypassed by an active adversary by simply stripping
the DKIM headers.

\myparagraph{DMARC}~\cite{rfc7489} is a comprehensive system that allows an
email-originating organization to express domain-level policies for email
management. A policy can specify how emails should be validated and how
receivers should handle validation failures. Additionally, DMARC policies can be
used to implement a reporting system (i.e., to report on actions performed under
a policy).  DMARC deploys SPF and DKIM, and domain owners can specify which of
those mechanisms (or both) should be used to validate their emails.  DMARC uses
\texttt{TXT} RRs to store policies, and the RRs are associated with domain names
prepended with the \texttt{\_dmarc.} prefix, e.g., \texttt{\_dmarc.a.com}.

\subsection{TLS PKI Enhancement}

\myparagraph{DANE}~\cite{rfc6698} allows domains to specify their key(s) or
key(s) of Certificate Authorities (CAs) they trust.  To this end, a domain
publishes a special DNS entry with its public key(s) or public key(s) of trusted
CA(s).  DANE introduces a new \texttt{TLSA} RR.   The scheme relies on DNSSEC,
requiring that the RRs be signed with the domain's DNSSEC key.  DANE records are
created per service, thus a DANE query encodes a transport protocol, and a port
number used. For instance, keys of a HTTPS server running at \texttt{www.a.com}
can be checked by querying \texttt{\_443.\_tcp.www.a.com}.  Such a flexible
mechanism allows to use DANE for all services that deploy TLS.

\myparagraph{CAA}~\cite{rfc6844} aims to provide trust agility and remove a
single point of failure from the TLS PKI. Specifically, it allows a domain to
specify (in DNS) CA(s) authorized to issue certificates for the domain.  This simple
procedure can prevent the two following threats:
\begin{enumerate*}[label=(\roman*)]
    \item compromised CA: a CA that is not listed by a domain cannot issue a
        valid certificate for the domain,
    \item identity spoofing: a benign CA can refuse certificate issuance if it is
        not listed by the domain.
\end{enumerate*}
CAA introduces new \texttt{CAA} RRs, which do not have to be protected via
DNSSEC, although it is recommended.

\myparagraph{Log-based approaches} are recent PKI enhancements that introduce
publicly-verifiable logs. The most prominent example is CT~\cite{rfc6962}, whose
goal is to make all certificates issued by CAs visible.  To this end, every
certificate is submitted to a log, which returns a signed \textit{promise} that
the certificate will be logged.  Then, in every TLS connection a client receives
a certificate accompanied with the logging promise.  However, it is important to
verify whether the promise was met, and to do so the client has to obtain a
proof from the log that given certificate indeed was logged.  Laurie et al.
propose~\cite{laurie2017certificatve} that clients ask a special CT-supported
DNS server for such a proof.  An advantage of this scheme is that DNS requests
are sent via a local resolver, thus the CT DNS server (and the log) cannot
identify the client, but only his resolver (usually run by his ISP).

\section{Current State of Deployment}
\label{sec:state}
First, we investigate deployment characteristics of the enhancements.  In
particular, we focus on factors that can influence reliability of DNS as a
transport (i.e., RRs used and response sizes).  To this end, we conduct a
measurement of the hundred thousand most popular domains of the Internet
(according to the Alexa list: \url{http://www.alexa.com/topsites}).
For each domain name we queried
for RRs that implement a given functionality.  We queried for DANE's RRs
specific to HTTPS, i.e., \texttt{\_443.\_tcp.}, and \texttt{\_443.\_tcp.www.}
prepended to a queried domain name.  We omitted log-based mechanism, as no
scheme is combined with DNS yet (up to our knowledge).
\vspace{-0.5cm}
\begin{table}[h]
\setlength\tabcolsep{4pt}
 \begin{center}
   \begin{tabular}{  l | l | r | r r r r }
     & \textbf{RR(s)} & \textbf{Successful} &
     \multicolumn{4}{c}{\textbf{Response Size (B)}}\\ 
     \textbf{Mechanism} & \textbf{Queried} & \textbf{Responses} &
     \textbf{min}&\textbf{med}&\textbf{avg}&\textbf{max}\\ \hline
     SPF & \texttt{TXT} & 53365 (53.37\%)     & 25  & 148  & 185 & 3138 \\
     & \texttt{SPF} & 4182 \ \ (4.18\%)           & 27  & 122  & 144 & 1606 \\ \hline
     Sender ID & \texttt{TXT} & 1766 \ \ (1.77\%) & 56  & 303  & 333 & 1285 \\
     & \texttt{SPF} & 98 \ \ (0.10\%)             & 79  & 234  & 247 &  538 \\ \hline
     DKIM & \texttt{TXT} & 5049 \ \ (5.05\%)      & 49  &  64  &  97 & 1007 \\ \hline
     DMARC & \texttt{TXT} & 7361 \ \ (7.36\%)     & 35  & 133  & 140 & 1003 \\ \hline
     DANE & \texttt{TLSA} & 48 \ \ (0.05\%)       & 80  &  88  &  96 &  182 \\ \hline
     CAA & \texttt{CAA} & 15 \ \ (0.02\%)         & 58  & 106  & 106 &  269
   \end{tabular}
     \caption{Measured scale of deployment and response sizes.}
\vspace{-2.0cm}
   \label{tab:responses}
 \end{center}
\end{table}
\begin{figure}[h]
  \centering
  \includegraphics[width=0.7\linewidth]{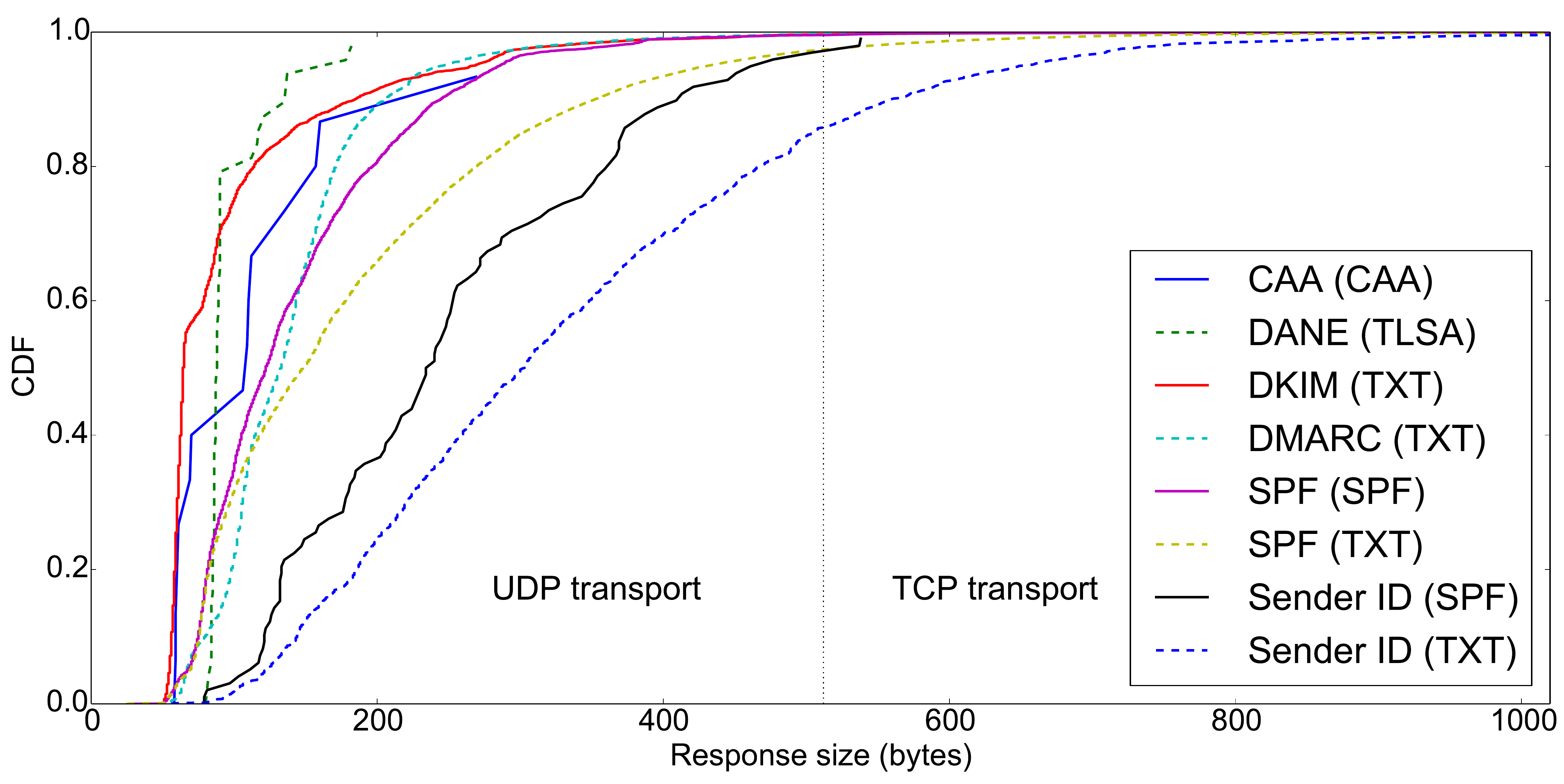}
\vspace{-0.4cm}
  \caption{CDF of the measured response sizes.}
\vspace{-0.5cm}
  \label{fig:respsize_cdf}
\end{figure}

\autoref{tab:responses} presents the measured scale of deployment with the
response size characteristics, while \autoref{fig:respsize_cdf} presents a CDF
of the measured response sizes.  As depicted, \texttt{TXT} RRs dominate,
constituting about 94\% of all successful responses. It is mainly due to
well-established deployment of the mail enhancements (SPF mainly). Although, new
RR types (like \texttt{SPF}) were introduced, the operators clearly prefer to
rely on older \texttt{TXT} RRs.  PKI enhancements do not have significant
deployment, which is probably caused by their relative immaturity (e.g., SPF was
introduced in 2006, while DANE and CAA in 2012 and 2013, respectively).  Another
finding is that most of the responses fit the limit of 512 bytes.  An exception
are responses including Sender ID's data (approximately $15\%$ of them exceed
the limit).

\label{sec:reliability}
\section{Reliability of DNS}
To investigate how reliable DNS is for the security enhancements, we conducted a
series of experiments using RIPE Atlas (\url{https://atlas.ripe.net/}), the
largest publicly available global testbed for network measurements.  RIPE Atlas
is a network of hardware devices, called \textit{probes}, used for active
Internet measurements.  It supports DNS measurements, and provides good
geographic coverage~\cite{bajpai2015lessons}.  Through the measurements we
wanted to answer the two following questions:
\vspace{-7pt}
\begin{enumerate}
    \item Are \texttt{TXT} RRs discriminated (dropped or manipulated) by
        some DNS clients/resolvers or network devices?
    \item How reliable is DNS in transporting UDP responses larger than 512
        bytes?
\end{enumerate}
\vspace{-7pt}
The first question is motivated by the importance of TXT RRs (see
\S\ref{sec:enhancements}) and by the common belief that a significant fraction of
\texttt{TXT} RRs is not transported correctly (probably due to its non-standard
type).  We investigate the second question to verify how the 512 bytes limit
for UDP DNS responses is enforced by the DNS infrastructure. This question
is important as the previous work indicates that the TCP support at DNS
resolvers is incomplete~\cite{huston2013question}, thus it is risky to rely on
the failover mechanism.  (Note, that RIPE Atlas does not expose an option to
check whether a probe's DNS client/resolver correctly handles responses with the
\texttt{truncated} flag set.)

In order to conduct the measurements, we launched an authoritative DNS server,
and prepared it with DNS responses of the following sizes:
\vspace{-7pt}
\begin{description}
    \item[494 bytes]: the size is below the 512 bytes limit, but it can handle
        most of the current responses (see \autoref{fig:respsize_cdf}).  We
        investigated transport over \texttt{A} and \texttt{TXT} RRs, to verify
        whether \texttt{TXT} RRs are discriminated (while compared to \texttt{A}
        RRs).
    \item[1005 bytes]: responses with this size allow us to investigate how
        robust the DNS infrastructure is, when the UDP response size limit is
        exceeded. This size is also below the standard MTUs (i.e., about 1500
        bytes). 
    \item[1997 bytes]: by responses with this size, we want to investigate how
        exceeding the standard MTU influences DNS transport. 
\end{description}
\vspace{-7pt}

Our DNS server was configured not to set the \texttt{truncated} flag, and in the
RIPE Atlas setting we set the acceptable response size to 4096 bytes.  We scheduled
measurements on the RIPE Atlas at the end of August 2016. We assigned all 9270
connected probes to query our DNS server.  For response sizes of 1005 and 1997
bytes we investigated only \texttt{TXT} RRs.  Depending on the queried target,
the following number of probes have responded: 8952 for queried \texttt{A} and
\texttt{TXT} RRs sent in 494 bytes responses, 8934 for 1005 bytes responses, and
7990 for 1997 bytes responses. Note, that each probe could respond with multiple
DNS responses.

In \autoref{tab:res} we present the obtained results.  As probes can use the
same, popular resolvers, beside the absolute number of responses, we also
present results for unique resolutions, where a unique resolution is defined as
a triple: number of RRs within a response, response size, and resolver's
address.
\begin{table}[t]
    \footnotesize
        \setlength\tabcolsep{4pt}
 \begin{center}
   \begin{tabular}{l | c | r| r r r | r r r r }
       \multicolumn{3}{c|}{}
     & \multicolumn{3}{c|}{\textbf{Successful Resolutions}} &
     \multicolumn{4}{c}{\textbf{Failed Resolutions}}\\ 
     & \textbf{Test} & \textbf{Total}& \textbf{Total} &
       \textbf{Exact} & \textbf{Larger} & \textbf{Total} &
       \textbf{Error}&\textbf{Empty}&\textbf{Truncated}\\ \hline
\multirow{8}{*}{\rotatebox{90}{\textbf{All Responses}}}
& \texttt{A}  & 16570 & 15468 & 15356 & 112 & 1102 & 867 & 189 & 46 \\
& 494B    & 100\% & 93.35\% & 92.67\% & 0.68\% & 6.65\% & 5.23\% & 1.14\% &
     0.28\% \\ \cline{2-10}
& \texttt{TXT} & 16570 & 15460 & 15343 & 117 & 1110 & 892 & 206 & 12 \\
& 494B & 100\% & 93.30\% & 92.60\% & 0.71\% & 6.70\% & 5.38\% & 1.24\% & 0.07\% \\ \cline{2-10} 
& \texttt{TXT}& 16553 & 13480 & 936 & 12544 & 3073 & 1504 & 1155 & 414 \\
& 1005B   & 100\% & 81.44\% & 5.65\% & 75.78\% & 18.56\% & 9.09\% & 6.98\% & 2.50\% \\ \cline{2-10} 
& \texttt{TXT}& 13727 & 7286 & 29 & 7257 & 6441 & 2360 & 3617 & 464 \\
& 1997B   & 100\% & 53.08\% & 0.21\% & 52.87\% & 46.92\% & 17.19\% & 26.35\% & 3.38\% \\
\hline\hline
\multirow{8}{*}{\rotatebox{90}{\textbf{Unique Responses}}}
& \texttt{A} & 7452 & 6625 & 6526 & 99 & 827 & 633 & 166 & 28 \\
& 494B     & 100\% & 88.90\% & 87.57\% & 1.33\% & 11.10\% & 8.49\% & 2.23\% & 0.38\% \\ \cline{2-10} 
& \texttt{TXT} & 7447 & 6618 & 6516 & 102 & 829 & 638 & 181 & 10 \\
& 494B & 100\% & 88.87\% & 87.50\% & 1.37\% & 11.13\% & 8.57\% & 2.43\% & 0.13\% \\ \cline{2-10} 
& \texttt{TXT} & 7938 & 6222 & 450 & 5772 & 1716 & 922 & 636 & 158 \\
& 1005B    & 100\% & 78.38\% & 5.67\% & 72.71\% & 21.62\% & 11.62\% & 8.01\% & 1.99\% \\ \cline{2-10} 
& \texttt{TXT} & 6887 & 3741 & 19 & 3722 & 3146 & 1252 & 1652 & 242 \\
& 1997B    & 100\% & 54.32\% & 0.28\% & 54.04\% & 45.68\% & 18.18\% & 23.99\% & 3.51\% \\
   \end{tabular}
   \caption{Measured reliability of DNS.}
\vspace{-1.5cm}
   \label{tab:res}
   \label{tab:res_uniq}
 \end{center}
\end{table}
The successful results are divided into responses that were received with the
exact size served (by the authoritative DNS server), and larger responses
(resolvers add other information that is relevant to the query, like addresses of
authoritative servers).  Failed resolutions are divided into three categories.
First, the fraction of resolution errors is presented. These are errors such as
a DNS resolver that could not be found, or a failed connection. Then, we present
empty DNS responses (i.e., number of answers equals zero). The last category
shows the number of truncated responses, i.e., responses with fewer number of
RRs than expected or/and shorter payload of the response.

Our first observation is that for the 494 bytes long responses
there is only a negligible difference between reliability of \texttt{A}-only
responses versus \texttt{TXT}-only responses.
Secondly, the results show that UDP responses with size above the 512 bytes
limit increase the failure rate from 6.70\% to 18.56\% (all responses) and from
11.13\% to 21.62\% (unique responses).  Taking into consideration the results
about failing TCP support, it might be more effective to use UDP with increased
size instead of TCP.
Lastly, the largest responses investigated (1997 bytes) are successfully
delivered only in about 50\% of all cases.  That is probably caused by MTU
issues, as common MTUs over the Internet are about 1500 bytes.
We also observe, that resolvers enlarge responses usually when they are large
already.

Although RIPE Atlas is an ideal open testbed for such tests, it introduces some
biases.  Probes are plug-and-forget devices, thus an owner may be not aware that
DNS resolution at his/her probe does not work properly (this could explain the
large fraction of DNS errors even for the smallest responses investigated).
Moreover, probes are usually installed by network-savvy users like research
institutions, Internet operators, hobbyists, and the probe distribution (based
on their ASes) is heavy-tailed~\cite{bajpai2015lessons}.

\section{Tor and Security Enhancements}
Tor~\cite{dingledine2004tor} is the most popular software and infrastructure for
enabling anonymous communication over the Internet. It is an onion routing
protocol, where an \textit{encryption circuit} is selected by the Tor client
software. DNS querying over Tor is also anonymous and conducted by an
\textit{exit node} of the circuit (this node will forward traffic to
destinations). 

The DNS resolution in Tor is restricted only to \texttt{A}, \texttt{AAAA}, and
\texttt{PTR} RRs. This obviously limits the deployment of DNS-supported security
enhancements in Tor.  It is especially important for the PKI enhancements, as
they assume clients to participate in the protocol (the mail enhancements are
deployed mainly by the mail infrastructure).  

In this section, we investigate whether the supported RRs can be used to
implement DNS-supported enhancements (for instance, one could convey information
on a series of \texttt{A} or \texttt{AAAA} RRs).
\begin{wrapfigure}{R}{0.45\textwidth}
    \vspace{-.80cm}
  \begin{minipage}[b]{0.45\textwidth}
    \centering
    \includegraphics[width=\textwidth]{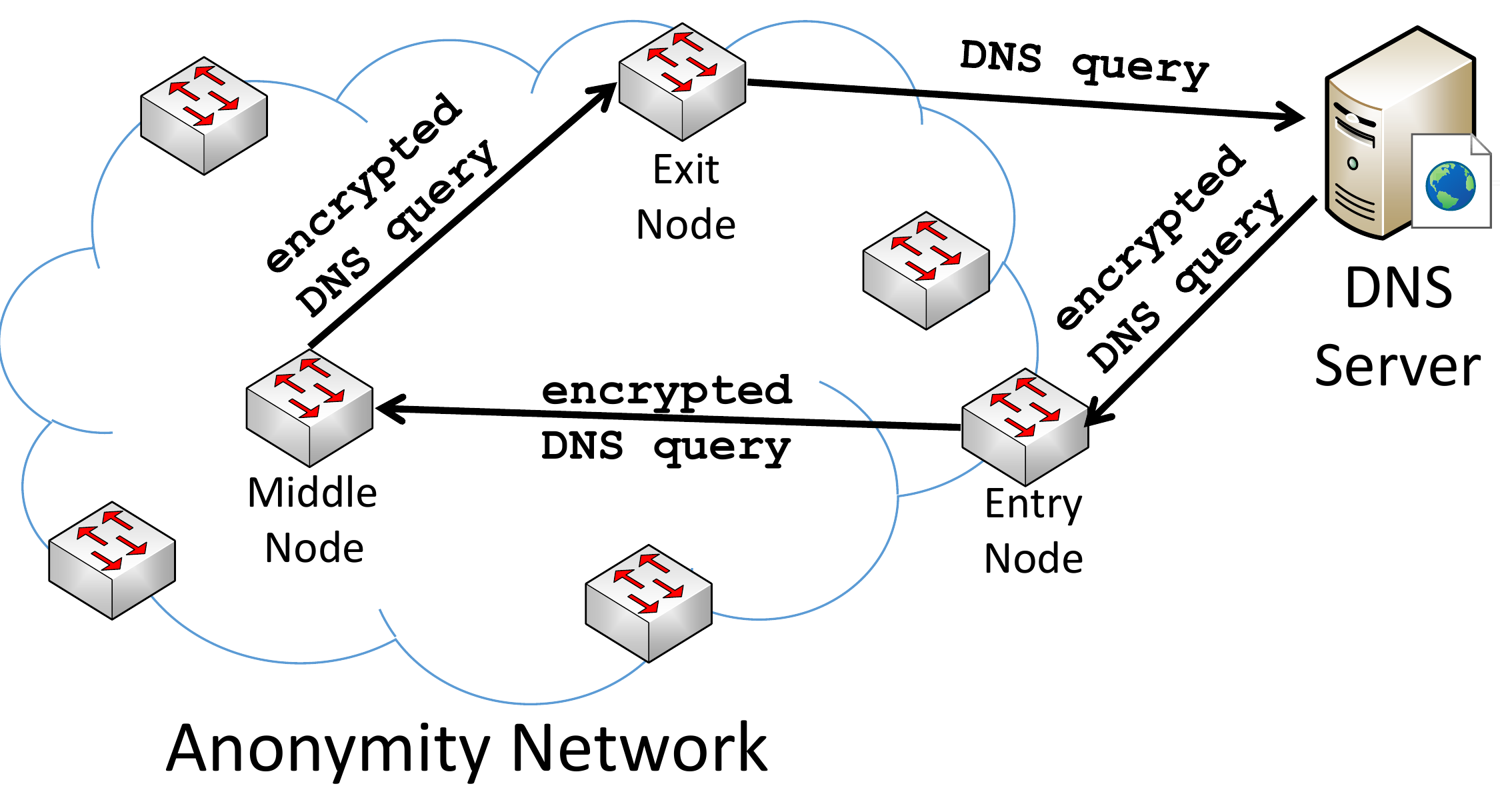}
    \vspace{-0.75cm}
    \captionof{figure}{Tor-based measurement scenario.}
    \label{fig:tor}
    \centering
    \setlength\tabcolsep{2pt}
 \begin{center}
       \footnotesize
   \begin{tabular}{ c | c | c | c }
       \texttt{A}&\texttt{PTR} (IPv4)&\texttt{PTR} (IPv6)&\texttt{AAAA}\\ \hline
       99.78\% & 99.22\% & 98.89\% & 23.05\%
   \end{tabular}
    \vspace{-0.25cm}
     \captionof{table}{Fraction of successful resolutions (i.e., single RR
     returned) depending on type.}
   \label{tab:tor1}
 \end{center}
    \end{minipage}
    \vspace{-1.1cm}
\end{wrapfigure}
We measured DNS resolution over Tor, using our authoritative server, that was
also configured as a Tor Linux client (i.e., the server queried itself through the Tor
network, as presented in \autoref{fig:tor}).  For every set of queries, a new
Tor circuit was selected, and we conducted 15000 such resolutions.
We investigated how reliable Tor is in resolving requests for the supported RRs
(i.e., \texttt{A}, \texttt{AAAA}, and \texttt{PTR}). We checked \texttt{PTR}
queries for both, IPv4 and IPv6 addresses.    

The first observation is that all asked resolvers limited DNS responses only to
a single RR.  This limits ways the supported RRs can be used to encode some
additional data (e.g., single \texttt{A} query can return only four bytes).
Table~\ref{tab:tor1} presents the fraction of successfully resolved requests.
As depicted, \texttt{A} queries are resolved slightly more reliably than
\texttt{PTR} queries for IPv4 addresses, which in turn are less reliable for
IPv6 addresses.    The results also show, that although \texttt{AAAA} RRs are
supported, they are resolved correctly only for 23\% of requests (probably, only
nodes supporting IPv6 resolve them).

\begin{table}[b]
       \vspace{-0.5cm}
 \begin{center}
    \setlength\tabcolsep{3.5pt}
   \begin{tabular}{ c | c | c | c | c | c | c | c | c | c}
       \textbf{61B} & \textbf{110B} & \textbf{158B} & \textbf{254B} &
       \textbf{366B} & \textbf{398B} & \textbf{430B} & \textbf{462B} &
       \textbf{478B}& \textbf{494B}\\
       1 RRs & 4 RRs & 7 RRs & 13 RRs & 20 RRs & 22 RRs & 24 RRs &
       26 RRs & 27 RRs & 28 RRs\\
       \hline
       99.77\% & 99.77\% & 99.77\% & 99.77\% & 99.23\% & 99.16\% & 98.10\% &
       92.87\% & 91.27\% & 38.36\%\\
   \end{tabular}
     \caption{Fraction of successful resolutions (i.e., single \texttt{A} RR
     returned) depending on the response size (from the authoritative server).}
   \label{tab:fail_size}
 \end{center}
\end{table}
Surprisingly, we observed that some resolvers fail to return any response when
the response from the authoritative server is large (but still below 512 bytes).
To further investigate this phenomena, we prepared responses with \texttt{A} RRs
with different sizes.  We then  measured when requests are processed successfully
(by success we mean a response to the client that contains a single RR, although
many were served). The results (see \autoref{tab:fail_size}) show that
reliability of DNS resolution decreases with the response size.  Only
38\% of all resolutions succeeded at all with 494 bytes long responses served.

\section{Conclusions}
\label{sec:conclusions}
Our study confirms that DNS-based security enhancements should respect the
conservative limit of 512 bytes for responses, as robustness of DNS transport
can be influenced by many uncontrollable factors.  Fortunately, the limit is
sufficient for about 95\% of all received responses.  However, our study does
not confirm the common belief that \texttt{TXT} RRs are being discriminated.
Our work identifies DNS resolution in Tor as an interesting subject for future
work, as we found it surprising and inconsistent: resolvers fail to return large
responses, slightly differently handle \texttt{PTR} RRs for IPv4 and IPv6
addresses, \texttt{AAAA} RRs are officially supported, but in practice are
resolved only by $23\%$ of all resolvers.  We also observe that restricting
other RRs (especially PKI-related, like \texttt{TLSA}) will actually decrease
security of end users. Hence, to fulfill Tor's mission (i.e., \textit{``to allow
people to improve their privacy and security on the Internet''}) the developers
should consider supporting DNS-based security enhancements.

\section*{Acknowledgment}
We gratefully acknowledge support from ETH Zurich and from the Zurich
Information Security and Privacy Center (ZISC).
We thank Brian Trammell and the anonymous reviewers, whose feedback helped to
improve the paper.

\bibliographystyle{abbrv}
\bibliography{ref,rfc}
\end{document}